\documentclass[useAMS,usenatbib]{mnras}

\usepackage{color} 
\usepackage{soul} 
\usepackage[T1]{fontenc}
\usepackage{ae,aecompl}
\usepackage{psfig}
\usepackage[dvips]{graphicx}
\usepackage{lscape}
\usepackage{amssymb,amsmath}     
\usepackage{color}                        

\def\lsim{\mathrel{\hbox{\rlap{\hbox{\lower4pt\hbox{$\sim$}}}\hbox{$<$}}}}
\def\gsim{\mathrel{\hbox{\rlap{\hbox{\lower4pt\hbox{$\sim$}}}\hbox{$>$}}}}

\title{Is ram-pressure stripping an efficient mechanism to remove gas in galaxies?}

\author[V. Quilis et al.]{
 Vicent Quilis$^{1,2}$\thanks{e-mail: vicent.quilis@uv.es}, Susana Planelles$^{1}$, 
 Elena Ricciardelli$^{3}$\\
$^{1}$Departament d'Astronomia i Astrof\'{\i}sica, Universitat de Val\`encia,
c/ Dr. Moliner 50, E-46100 - Burjassot, Val\`encia, Spain\\
$^2$Observatori Astron\`omic, Universitat de Val\`encia, E-46980 Paterna, Val\`encia, Spain\\
$^{3}$Laboratoire d'Astrophysique, \'Ecole Polytechnique F\'ed\'erale de Lausanne (EPFL), 1290 Sauverny, Switzerland }

\begin{document}
\date{Accepted ...  Received ...; in original form ...
}
\pagerange{\pageref{firstpage}--\pageref{lastpage}} \pubyear{...}
\maketitle

\label{firstpage}

\begin{abstract}

We study how the gas in a sample of galaxies ($M_*>10^9 \,M_{\odot}$)  in  clusters,  obtained in a cosmological simulation,  
is affected by the interaction with the intra-cluster medium (ICM). 
The dynamical state of each elemental parcel of gas is studied using the total energy.   
At  $z\sim 2$, the galaxies in the simulation are evenly distributed within clusters, moving later on towards more central locations. 
In this process, gas from the ICM is accreted and mixed with the gas in the galactic halo. Simultaneously, the interaction with the environment removes part of the gas.
A characteristic stellar mass around $M_* \sim10^{10}M_{\odot}$ appears as a threshold marking two differentiated behaviours. Below this mass,  galaxies are located at the external part of clusters and have eccentric orbits. The effect of the interaction with the environment is marginal. Above, galaxies  are mainly located at the inner part of clusters with mostly radial orbits with low velocities. 
In these massive systems, part of the gas, strongly correlated with the stellar mass of the galaxy, is removed. The amount of removed gas is subdominant compared with the quantity of retained gas which is continuously influenced by the hot gas coming from the ICM. 
The analysis of individual galaxies reveals the existence of a complex pattern of flows, turbulence and a constant fueling of gas to the hot corona from the ICM that  could make the global effect of the interaction of galaxies with their environment to be substantially less dramatic than previously expected.

\end{abstract}

\begin{keywords}
galaxies: haloes --- galaxies: formation --- galaxies: evolution --- galaxies: ISM --- galaxies: clusters: intra-cluster medium --- galaxies: interactions --- methods: numerical
\end{keywords}

\section{Introduction}

Galaxies living in dense regions of the Universe suffer dramatic transformations throughout their existence due to the interaction with the environment. Other galaxies, and galaxy clusters themselves, produce tidal interactions resulting in the removal of part of the gas  content of galaxies \citep{1996Natur.379..613M, 2003ApJ...582..141G,2007ApJ...671.1434T}. Although tidal interactions can be important, the motion of galaxies through the intra-cluster medium (ICM) produces an additional pressure that can efficiently strip away a large fraction of the interstellar medium (ISM) gas in galaxies. 
This mechanism, known as ram-pressure stripping (RPS), was first explained by \cite{1972ApJ...176....1G} as a simple balance between the pressure gradient and gravity, but it turned out to be much more complex due to the role of turbulence and viscous effects  \citep[e.g.][]{Roediger_2008}. 

Numerical simulations have played a crucial role to understand the RPS effects on galaxies. Many groups have 
carried out simulations to study the effects of RPS depending on the morphology of galaxies (spiral vs elliptical), the masses of galaxies (dwarfs), the orientation of the orbits, the density of the environment, or the correlation with the star formation, among others 
\citep[e.g.][]{1999MNRAS.308..947A, 2000Sci...288.1617Q, 2001ApJ...561..708V, 2001MNRAS.328..185S, 2006MNRAS.371..609R,
2006MNRAS.369.1021M, 2007A&A...472....5J, 2008A&A...481..337K, 2009ApJ...694..789T, 2009A&A...499...87K}.
RPS has also been a very appealing topic from the observational point of view since there are plenty of observational evidences of its relevance. Thus, it has been observed in galaxies in the Virgo cluster, both spirals \citep[e.g.][]{1990AJ....100..604C,
2004AJ....127.3361K, 2005AJ....130...65C} and ellipticals \citep[][]{2006ApJ...644..155M}.  RPS has been also reported in galaxies in Coma and other clusters  \citep[see, for instance,][]{2000AJ....119..580B, 2005ApJ...624..680K, 2007AJ....133.1104L, 2007MNRAS.376..157C,
2011AJ....141..164A, 2014ApJ...781L..40E} as tails of atomic gas (HI), as ionized gas detected in H$\alpha$ \citep[e.g.][]{2007ApJ...660.1209Y,2013ApJ...777..122Z}, or, even in some cases, stripped gas can be seen as molecular clouds \citep{2014ApJ...792...11J}. 

The gas content in galaxies can be separated in two thermodynamically well different components. There is a cold gas component that forms part of the ISM, mainly located  in the discs and in the central parts of galaxies. This cold component, directly connected with the star formation in galaxies, is effectively removed by RPS in short time scales \citep{2000Sci...288.1617Q}. In addition to the cold gas, it is known that galaxies are surrounded by a  hot ($T\sim 10^6 \, K$) extended halo. This hot corona is a source of diffuse soft X-ray emission that has been observed around both elliptical \citep[e.g.][]{1985ApJ...293..102F, 2009ApJ...693.1142S} and spiral galaxies  \citep[e.g.][]{2004ApJS..151..193S,2013MNRAS.428.2085L}. Simulations have shown that this hot corona can also be dramatically affected by RPS \citep[e.g.][]{2008ApJ...672L.103K, 2008MNRAS.383..593M,2015MNRAS.449.2312V}. This process would not have a direct impact on the star formation in short time scales, as the removal of cold gas but, it would translate into a reduction of gas able to cool  and,  therefore, to fuel the disc. The effect of this lack of cold gas would produce a fading away of the star formation over time scales of several Gyr. This process is commonly known as `starvation' or `strangulation'. 

The expected consequence of the stripping of the hot halo would  be a deficiency of  X-ray emission of galaxies in clusters when compared with similar mass systems in the field.  However, {\it Chandra} observations of different morphological galaxy types have shown that the presence  of a hot galactic extended corona is ubiquitous \citep{2007ApJ...671..190S,2008ApJ...679.1162J} and, therefore, in apparent conflict with what expected. 
\cite{2010ApJ...715L...1M} suggested that very hot gas ($T\sim 10^8 \, K$) in the ICM could confine the relatively cold gas of the corona ($T\sim 10^6 \, K$) due to the pressure gradient and, therefore, produce a competing effect against the RPS. This scenario has been numerically explored by \cite{2012MNRAS.424.1179B}.

Most of the theoretical works previously quoted  analyze, however,  the effects of RPS on idealized individual galaxies suffering a wind. In other few cases, simplified galaxies  are distributed throughout groups and galaxy clusters set up according to  analytical profiles. In the present paper, we study the effects of RPS  on a sample of galaxies obtained in a fully cosmological simulation without any  simplification or assumption on the clusters or on the galaxies in clusters. 
This simulation provides us with a full description of all the components of galaxies along the cosmic time, being possible to track in detail every single galaxy and the evolution of its gas content.  The effects of the RPS on both the hot (corona) and the cold gas can be addressed by following the evolution of both components.

The paper is structured as follows.   In Section 2, we describe the numerical methodology that has been used to produce the cosmological simulation,  and the techniques to analyze it.  The results of the analysis of the simulation on the fate of the gas in the galaxy haloes 
are presented in Section 3. Finally, we discuss the main results of this work and the conclusions in Section 4.

\section{Numerics}

\subsection{The simulation}
In  this  paper we analyse a simulation performed  with  the
cosmological  code  MASCLET \citep{2004MNRAS.352.1426Q}. 
MASCLET is an Eulerian code,  based  on  {\it high-resolution  shock  capturing}
techniques  to evolve the gaseous (collisional) component,  coupled with an N-body approach 
(multigrid particle mesh) to describe  the collisionless component (dark matter).  
The gravity solver couples the evolution of both components.
In order to improve the spatial and temporal resolution, MASCLET is based on 
an  adaptive  mesh refinement  (AMR) scheme.

The simulation assumes a spatially  flat $\Lambda
CDM$  cosmology, with the following  cosmological  parameters:  $\Omega_m=0.31$,
$\Omega_{\Lambda}=\Lambda/{3H_o^2}=0.69$, $\Omega_b=0.048$, 
$h=H_o/100  km\, s^{-1}\,Mpc^{-1}=0.678$,  $n_s=0.96$ and $\sigma_8=0.82$.

The simulated region is discretised  with  $128^3$ cubical cells within a cube of comoving  side length $40\,  Mpc$. 
By employing a  CDM transfer function from \cite{1998ApJ...496..605E}, we  set up the initial conditions at  $z=100$. 
We applied a constrained realization in order to generate a rich galaxy cluster in the centre of the box \citep[see][]{1991ApJ...380L...5H}.

From the initial conditions, evolved until present  time using a low resolution domain,
we select regions satisfying some refining criteria in order to arrange three  refinement levels 
($l=1, 2$, and $3$) for the  AMR scheme.
In  these initially refined  levels, the  dark matter (DM)  component is sampled with  DM  particles  
8, 64, and 128 times, respectively, lighter than  those used  to sample regions in the coarse grid ($l=0$).   
As the evolution proceeds, the local baryonic  and DM densities are used to refine regions on  the different grids.
In  our   AMR  scheme, the ratio between the cell sizes for a  given level ($l+1$)  and its
parent  level  ($l$)   is   $\Delta x_{l+1}/\Delta  x_{l}=1/2$, which is a  compromise between the
appearance of numerical instabilities and the increase in resolution.

In the present simulation we use a maximum of nine refinement levels
($l=9$), allowing for a peak physical spatial resolution of $\sim 610\,  pc$ at $z=0$. 
Four different particles species are considered for the DM,
corresponding to those particles on the coarse grid and those
within the three first  levels of refinement.  The
best  mass resolution  is  $\sim 2\times  10^6\,  M_\odot$, equivalent to use 
$1024^3$ particles in the whole computational domain.

Cooling  and heating processes, including inverse Compton and free-free  cooling, UV heating \citep{1996ApJ...461...20H} and
atomic and molecular cooling for a primordial gas, are implemented in our simulation.  
We  consider that the  gas is optically thin  and in ionization  equilibrium, but  not in  thermal equilibrium
\citep{1996ApJS..105...19K,1998MNRAS.301..478T}, in order to compute the abundances  of each species.  
We use the tabulated  cooling rates from \citet{1993ApJS...88..253S}, which depend on the local metallicity.
For  temperatures below $10^4\,K$ the cooling curve was  truncated.

\subsection{Star formation}
\label{starformation}
 
In MASCLET the star  formation is implemented according to \citet{1997MNRAS.284..235Y} 
and \citet{2003MNRAS.339..289S}. In  our particular scheme, we consider that cold  gas in a cell is transformed into
stellar  particles  according  to
$\dot{\rho_*}=-\dot{\rho}=(1-\beta)\,{\rho}/{t_*(\rho)}$,  where $\rho$
and  $\rho_*$  are the  gas  and  star  densities, respectively, and $t_*$ 
is a  characteristic time  scale.   The
parameter  $\beta$ (we use $\beta=0.1$, consistent with a Salpeter IMF)  represents the  mass  fraction  
of massive  stars ($>8\,M_{\odot}$) that explode as  supernovae (SNe) and return, therefore,  to
the  gas  component in  the  cells.  For the characteristic star formation
time scale,  we  also assume $t_*(\rho)=t^*_o(\rho/\rho_{_{th}})^{-1/2}$,  consistent with
$\dot\rho_* \sim \rho^{1.5}/t^*_o$   \citep{1998ApJ...498..541K}.   
In   this   manner,  there is a dependence on the local  dynamical time of the gas, 
the density threshold for star formation ($\rho_{_{th}}$),
and  the associated  characteristic time  scale ($t^*_0$).   In particular,  we  use 
$t^*_0=2\,  Gyr$  and $\rho_{_{th}}=10^{-25}\,g\,cm^{-3}$.  
As for the energy,  we assume that each  SN injects $10^{51}\, erg$ of thermal energy in  the original cell. 
Similarly, we also consider that every star formation episode returns to the medium a metal fraction given by
$y=\frac{m_{_M}}{m_*}$, where $y$ (we adopt $y=0.02$), $m_{_{M}}$, and $m_*$ are, respectively,  the yield, the mass of metals, and the 
stellar mass.
From the metal density $(\rho_{_{M}})$ we define a metallicity, $Z=\frac{\rho_{_{M}}}{\rho}$,
that can be employed to compute cooling rates. 
A continuity equation, similar to the one used for the gas component, is employed to advect the metallicity throughout the computational domain. 
We do not account for the feedback from stellar winds (AGB stars) nor from the SN type Ia .

In our implementation,  we consider that star formation takes place
once every global  time step, $\Delta t_{l=0}$, and  only in the cells
at the highest levels of refinement ($l=7,8,$ and $9$).  Those cells at these refinement levels,
with gas  temperature $T  < 2\times10^4\,    K$    and    gas    density   $\rho    >
\rho_{_{th}}=10^{-25}\,  g\,  cm^{-3}$,  are allowed  to  form
stars.   In  these  cells,  collisionless  star  particles  with  mass
$m_*=\dot\rho_*\Delta  t_{l=0}\Delta x_l^3$ are  formed.  An additional condition, to
avoid abrupt changes in the  local gas density, constrains
the   mass of the star particles  to  be    $m_*={\rm min}(m_*,\frac{2}{3}m_{gas})$, 
where  $m_{gas}$ is the  total gas mass in the considered cell.  
The energy associated to the stellar feedback from SNe is deposited
within the same cell where the stellar particle is formed. 

\subsection{Halo identification and galaxy sample}
\label{sample}

The outcome of  our simulation is a complete  description of the three
components included  in the calculation,  namely, gas, dark  matter and
stars.  In  order to  analyse and characterise  the properties  of the
different structures that can be found in the simulation, we use two
different but complementary numerical tools briefly described here:

\begin{itemize} 

\item The dark matter halo finder ASOHF \citep[see][for further details]{2010A&A...519A..94P, 2011MNRAS.415.2293K}. ASOHF is an adaptive spherical overdensity halo finder able to identify and characterise the sample of dark matter haloes formed along the evolution of the simulation. A catalogue with the main halo properties, namely, positions, virial masses and virial radii, is provided.
   
\item To identify stellar haloes (galaxies) in the simulation, we use our own implementation of a friends-of-friends algorithm \citep[see][for further details]{2013MNRAS.436.3507N}. This stellar halo finder uses an adaptive linking length together with a phase-space control to ensure that all the stellar particles geometrically linked to a stellar halo are, indeed, real halo members.  Only galaxies with a minimum of 25 bound stellar particles are considered.
Each numerical galaxy identified in our simulation is characterised by its position, its stellar mass and its stellar half-mass radius.  
    
\end{itemize}

The outcome of our simulation is a list of 81 temporal snapshots, ranging form $z=100$ down to $z=0$, providing a complete description of the formation and evolution of a sample of synthetic cosmic structures. The analysis of this virtual universe, by means of our dark matter and stellar halo finders, produces a catalogue of galaxies and galaxy clusters at all redshifts. In the study presented in this paper, we will focus on galaxies with stellar masses $M_*>10^9 \,M_{\odot}$, which are the best numerically resolved. Taking into account this constraint, at $z=0$ there are 20 galaxies with stellar masses larger than this threshold. Concerning galaxy clusters,  we consider as clusters all the dark matter haloes  with total virial masses greater than $10^{13}\, M_{\odot}$. Thus, at  $z=0$, a total of 8 of these objects can be found in the simulation. We note that  galaxies in our sample are mostly located within the clusters, corresponding to the most massive ones within each cluster. It has been checked that they are not in interaction and that they have not undergone any major major merger since $z\sim2$. 

It is a well-known issue that numerical simulations produce a lower number of massive subhaloes (galactic-size haloes) 
than observed \citep[see, for instance,][and references therein]{Munari_2016}.
In this regard, we note that the relatively low-number statistics of objects in our sample
is a direct consequence of the simulation properties, namely, a moderate-size full cosmological box without any kind of
zoom-in or re-simulation. Nevertheless, we would like to point out that our results, in terms of numbers of
dark matter and galactic haloes, are in line with previous numerical estimates 
\citep[e.g., see][]{Springel_2008, Contini_2012, Knebe_2013, Eagle_2016}.

\subsection{Status of the gaseous component of galaxies: bound or unbound gas}
\label{method}

The common procedure to quantify the relevance of the RPS has been to estimate the balance between the drag force exerted by the environment and the gravitational force produced by the galaxy.  In this approach, the following condition is required \citep{1972ApJ...176....1G}
\begin{equation}
\rho_{_{ICM}}v^2 > \Sigma_{_{ISM}}\frac{d\phi}{dz},
\label{rps_gg}
\end{equation}
where $\rho_{_{ICM}}$ is the intra-cluster medium density, $v$ is the relative velocity of the galaxy with respect to the ICM, $\Sigma_{_{ISM}}$ is the surface gas density of the galaxy, and $\frac{d\phi}{dz}$ stands for the gravitational pull along the direction of motion of the galaxy.  This method requires to use several integrated quantities along the direction of motion of the galaxy, together with a surface with a fix geometry wrapping the galaxy.

In the present paper, we describe the effects of RPS with  a complementary approach based on the dynamical properties of each elementary volume of gas within the galactic halo. In our method, for every cell in the computational domain, we compute the total energy of such parcel of gas as the sum of the kinetic plus gravitational energies, $E_{t}=E_K + E_g$. The kinetic energy of each parcel of gas  is computed using the relative velocity referred to the centre of mass velocity of the galaxy. For the gravitational energy,  the total potential created by all the components in the galaxy (gas, dark matter and stars) is used. 
In this method, it is crucial to define the radius of the galactic halo. We define such radius as twice the stellar half-mass radius as provided by the stellar halo finder (see Sec.~\ref{sample}).  This is a well motivated assumption both from the theoretical point of view \citep[see, for instance,][]{Guo_2011} and from observational data \citep[e.g.][]{Noordermeer_2005}.  For each galactic halo, we compute the total energy of all cells contained within the sphere defined by the galactic radius. 
With this procedure, all the elemental parcels of gas within a galactic halo can be grouped in two wide categories. Those cells having gas with negative total energy contain gas that has not been detached from the galaxy due to the interaction with the environment. We define such  gas component as bound. On the contrary, cells enclosing gas with positive  total energy correspond to gas that has been stripped from the galaxy and that, eventually, will be removed, representing therefore the unbound gas component of the galactic halo.  

In our method, we identify unbound gas with stripped gas. This is obviously an  approximation but well motivated, though. Once a parcel of gas is unbound, 
its dynamics is detached from its former host galaxy and,  therefore, it moves freely within the ICM. It could be argued that this parcel of gas could be trapped back by the galaxy in the next time steps, but this situation would require a sudden increase of the gravitational force  exerted by the galaxy. This scenario, plausible though, is very unlikely -- from the statistical point of view -- as the change in the potential field created by a galaxy along several numerical time steps is smooth.  

This approach, significantly different from previous RPS analyses, has several advantages when compared with previous RPS studies based on  Eq.~\ref{rps_gg}. These last methods use to assume an homogeneous environment with a common bulk relative velocity and some fix surface wrapping the galaxy. On the other hand, the method proposed in this paper is local, meaning that the energy condition can be applied to every numerical cell  in which the galaxy is resolved. In this manner, 
it is possible to estimate the state of each parcel of gas regardless of its position within the galactic halo or its velocity. As a consequence, more realistic and inhomogeneous environments, with changing densities and velocities, can be easily treated without any extra assumption.  Besides, the proposed method does not require any special surface wrapping the galaxy and it naturally accounts for  all the gas contained within the galactic halo. 

At this point, it could be argued that  the physical mechanism producing the gas removal remains uncertain. 
Nevertheless, such a physical mechanism can be constrained thanks to the particularities of our galaxy sample. In this sense, as previously discussed, the galaxies in our sample are mainly the most massive ones within clusters, and  they have not experienced any major merger event since $z\sim2$. Therefore, these objects have not suffered any dramatic change due to mergers or tidal interactions, and these processes would not be behind the gas removal in our simulation. 
On the other hand, supernovae feedback, the only feedback source we consider in our simulation, could also play a role. However, it is well known that, although this feedback  can be crucial in galaxies with  $M_* < 10^{9} \, M_{\odot}$ \citep[e.g.][]{Brooks_2007, Silk_2012},  its role is marginal in larger systems like those considered in the present analysis. Therefore, the gas in our sample of galaxies is very unlikely to be unbound by the energy released by supernovae. Related with the inefficient feedback processes included in the simulation, as no AGN feedback is considered, our simulation presents the well-known overcooling problem. This excess of cooling prevents thermal evaporation from removing gas from galaxies. Thus, although other physical mechanisms can have a marginal role, the dominant effect in the simulation analysed in the present paper is the interaction of the galaxies with the ICM environment. This is the reason why  we will assume that the RPS is the dominant mechanism responsible of the gas removal in our virtual galaxies.  

\begin{figure}
\includegraphics[scale=.45]{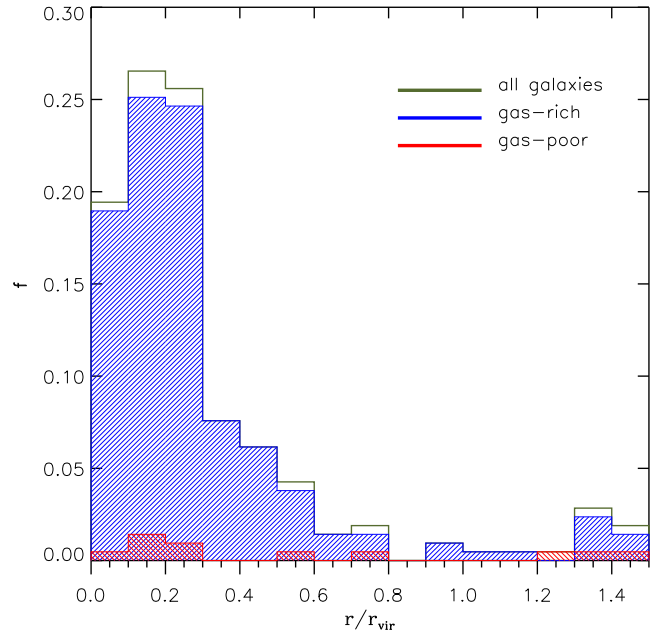}
\caption{Fraction of galaxies, $\rm f$, with respect to the total number of galaxies in the sample, as a function of their cluster-centric distance to their host galaxy cluster (in units of the cluster virial radius). 
We consider all galaxies in our simulation at $0<z<0.5$ and with stellar masses larger than $10^{9} \, M_{\odot}$. The green histogram stands for all the galaxies in the sample, while the blue (red) histogram represents the sample of gas-rich (-poor) galaxies.}
\label{fig1}
\end{figure}

\begin{figure}
\includegraphics[scale=.3]{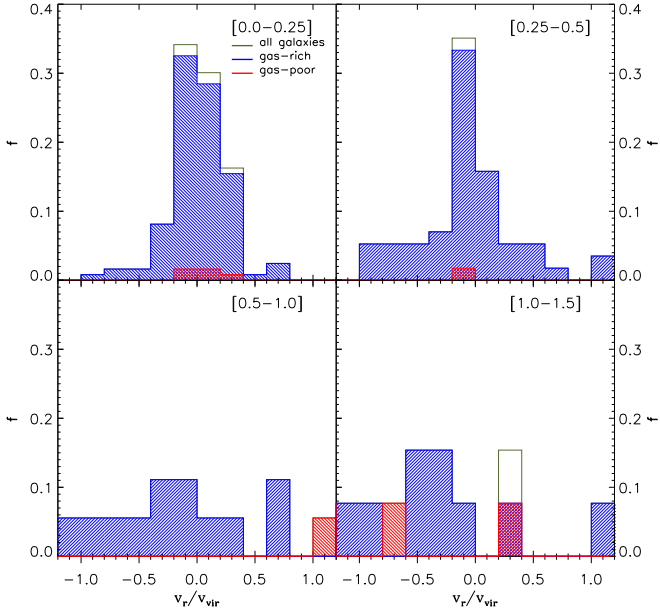}
\caption{Fraction of galaxies, $\rm f$, as a function of the ratio of the galaxies radial velocity to the virial velocity of their host galaxy cluster, i.e., $v_r/v_{vir}$. According to the radial distances from  each galaxy centre to the centre of its host cluster, our galaxy sample is divided in four cluster-centric distance ranges:  $r/r_{vir}<0.25$ (top-left panel), $0.25<r/r_{vir}<0.5$ (top-right),  $0.5<r/r_{vir}<1.0$ (bottom-left), and $1.0<r/r_{vir}<1.5$ (bottom-right).  We consider all galaxies with stellar masses larger than $10^{9} \, M_{\odot}$ and at $0<z<0.5$.  The fractions are computed with respect to the total number of galaxies within each distance range. Coloured histograms  have the same meaning than in Fig.~\ref{fig1}.}
\label{fig2}
\end{figure}

\begin{figure}
\includegraphics[scale=.3]{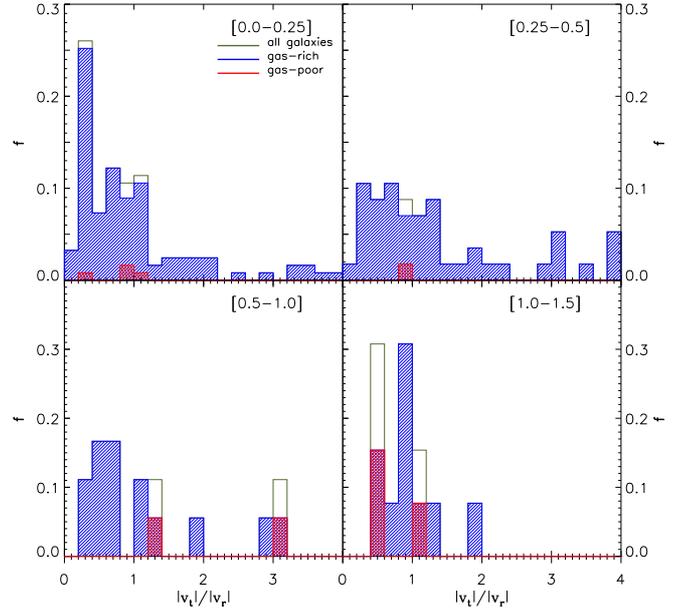}
\caption{Fraction of galaxies, $\rm f$, as a function of the ratio of the absolute values of their tangential to radial velocities.  Panels and colour histograms  have the same meaning than in Fig.~\ref{fig2}. }
\label{fig3}
\end{figure}

\section{Results}
\label{sec:results}

We use the galaxy  catalogue described in Sec.~\ref{sample} to study how the gaseous content of galaxies depends on the inherent galaxy properties but also on the interaction with the environment. 
To this end, we define the gaseous halo of a galaxy as the gaseous sphere centrered at the position of the  galaxy and with a radius twice the galaxy half-mass radius as provided by the halo finder. Within this sphere, we identify those gas parcels that are unbound, according to the criterion described in Sec.~\ref{method}. This procedure allows us to compute the amount of stripped gas (unbound) and the fraction of gas retained by the galaxy (bound). 

As a further step in our study, we classify our sample of galaxies in two different subsamples  depending on whether they are  gas-poor or gas-rich galaxies.  
The first subsample is made up by those galaxies with a large amount of gas compared to their stellar mass, that we take as all galaxies with a ratio $\frac{M_g}{M_*} > 0.1$. In the same manner, we define the subsample of gas-poor galaxies as those galaxies having values of  $\frac{M_g}{M_*} < 0.1$. This particular definition is a compromise compatible with both observations \citep[e.g.][]{2016arXiv161006932S} and simulations \citep[e.g.][]{Kauffmann_2016}.
 
Moreover, to have information on the spatial distribution of galaxies within clusters, all galaxies in our sample are assigned to their closest galaxy cluster. Then, for each galaxy, we compute the distance from the galaxy centre to the centre of the dark matter halo of the associated galaxy cluster. The centre and velocity of the galaxy clusters are defined as the centre of mass and the velocity of the centre of mass computed with all dark matter particles within the virial radius.

According to these criteria, Fig.~\ref{fig1} shows the fraction of galaxies (with respect to the total number of galaxies in the sample) as a function of $r/r_{vir}$, that is the ratio between the distance of each galaxy to the centre of its host cluster and the cluster virial radius. In this first analysis, we have considered all galaxies in our catalogue within a redshift range between $z=0.5$ and $z=0$. 
In addition,  depending on the amount of gas of the considered galaxies, we split the sample in gas-poor and gas-rich galaxies. From the analysis of this figure, there is a clear qualitative trend: at these low redshifts, the majority of galaxies in clusters are gas-rich and lay at distances below 0.8 times the virial radius of their host cluster, showing a maximum at around $r\sim 0.2 r_{vir}$. 
Although it has been reported that the fractions of gas-poor and gas-rich galaxies show some dependence on the mass of the host cluster, a lower number of gas-poor galaxies compared with the observations seems also to be a common trend in numerical simulations  \citep[e.g.][]{Kauffmann_2016}.

The radial distribution of our sample of galaxies is also in accordance with the results obtained from 
previous simulations, suggesting high stellar concentrations close to the centre of dark matter subhaloes \citep[e.g.][]{Nagai_2005}.

We note that, in this analysis, we have considered all galaxies in our sample within a redshift range between $z=0.5$ and $z=0$.
This approach allows us to increase the statistical relevance of our sample, although it implies that some objects can be included several times at different stages in their evolution.  In the following, this consideration will be applied when showing results at different redshift ranges. 

\begin{figure}
\centerline{\includegraphics[scale=.4]{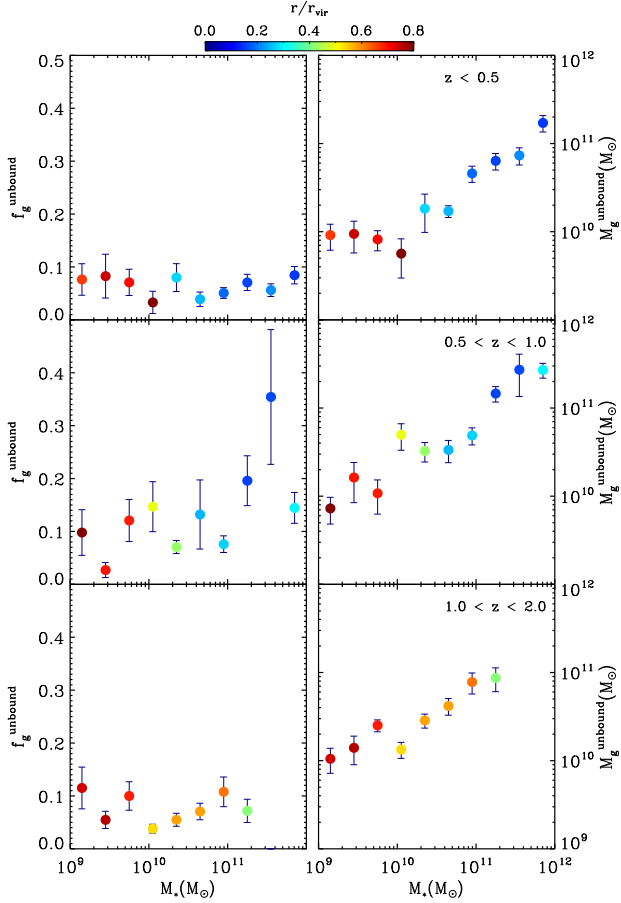}}
\caption{Mass of removable gas, $\rm M_g^{unbound}$ (right column), and mass fraction of unbound gas, ${\rm f_g^{unbound}}$ (left column), as a function of galactic stellar mass, $M_*$. Each row of panels represents all galaxies in the sample at three different redshift ranges. At a given epoch, galaxies are binned according to their stellar masses. For each mass bin, the average value of $\rm M_g^{unbound}$ and ${\rm f_g^{unbound}}$ together with the standard deviations  around the mean are represented by circles with error bars. The colour of each circle, according to the code displayed on top of the figure, represents the mean cluster-centric distance (in units of the virial radius) of all galaxies in a given mass bin.}
\label{fig4}
\end{figure}

\begin{figure}
\centerline{\includegraphics[scale=.7]{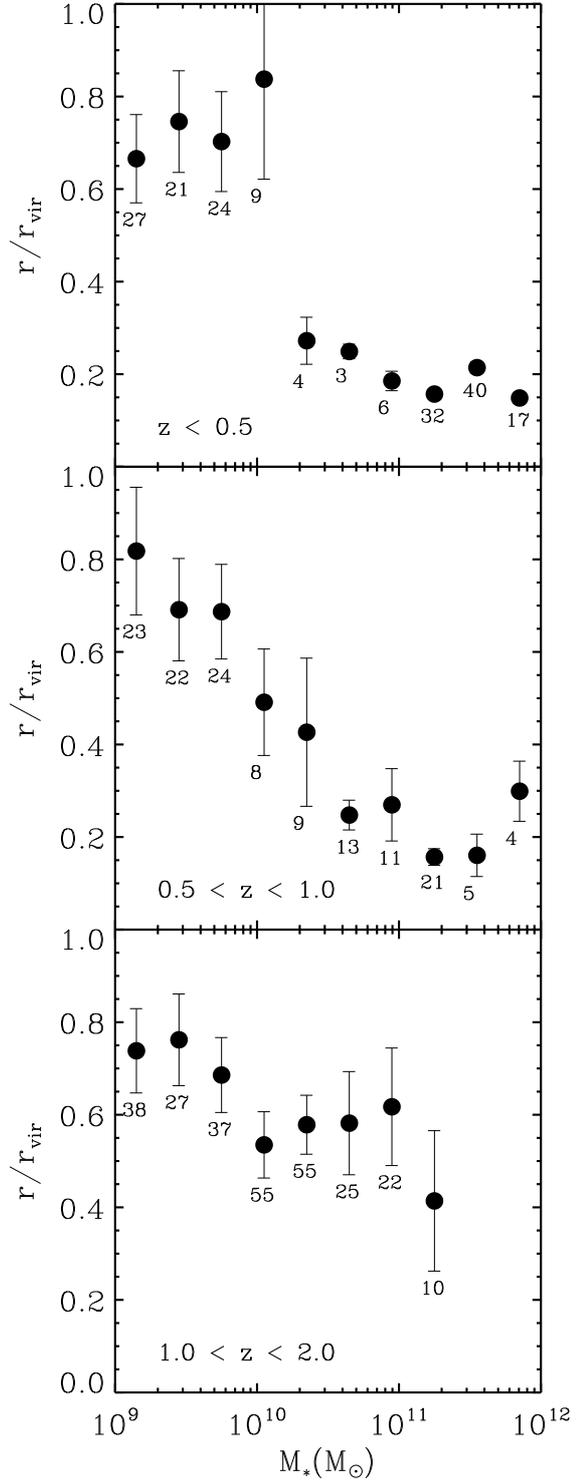}}
\caption{Average radial position of galaxies as a function of their stellar mass for three different redshift ranges: 
$1<z<2$ (bottom panel), $0.5<z<1$ (middle panel), and $z<0.5$ (top panel). The error bars stand for one standard  deviation. The number below each point displays the number of galaxies in each mass bin. The radial distances are in units of the virial radius of the galaxy cluster hosting each galaxy. } 
\label{fig5}
\end{figure}

\subsection{Effect of the motion of cluster galaxies on their gas content}

The motion of galaxies in the ICM is the physical mechanism responsible of removing part of their gaseous content through RPS. As a first step to deepen the effect of the RPS on the galactic gas content, we need to know how galaxies move inside their host galaxy clusters.
A relatively simple way to quantify this movement  is to study the gaseous content of galaxies depending on their velocities and positions within galaxy clusters.

\begin{figure}
\centerline{\includegraphics[scale=.4]{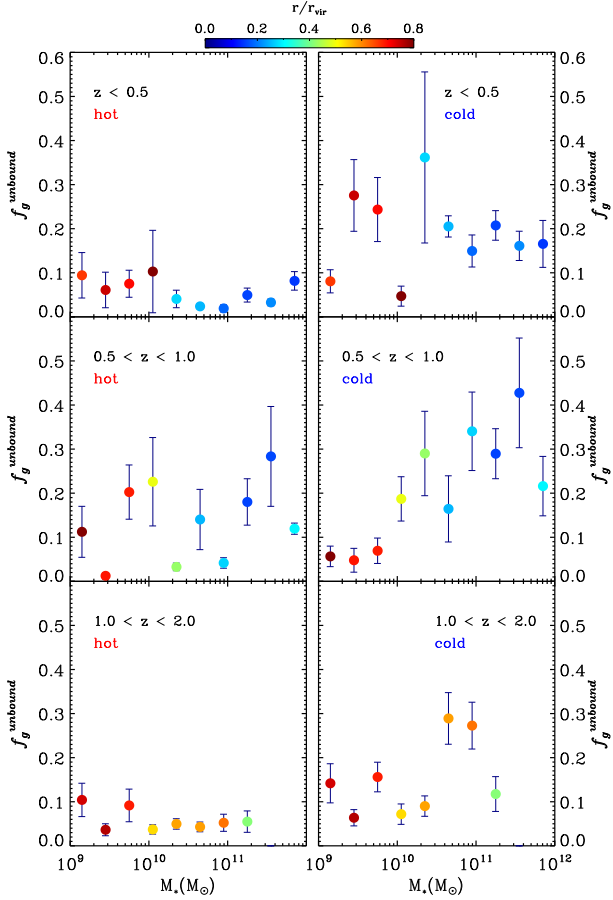}}
\caption{Redshift evolution of the relative fraction  of unbound gas ($\it f_g^{unbound}$), hot (T>$5\times10^4$ K; left column) and cold (T<$5\times10^4$ K; right column), as a function of galactic stellar mass, $M_*$. Points and colours have the same meaning than in  Fig.\ref{fig4}. To obtain this fraction, the amount of unbound cold (hot) gas is normalized over the total mass of the cold (hot) component. }
\label{fig6}
\end{figure}

Again, we consider all galaxies in our catalogue within a redshift range between $z=0.5$ and $z=0$ and compute the relative velocity of each galaxy to its host  cluster.
Once the relative velocities of the galaxies are known, we compute the radial ($v_r$) and tangential ($v_t$) velocity components. The comparison of these two components can shed light on the dynamical state of the galaxies within their host galaxy clusters. Thus, the ratio $v_r/v_{vir}$, where $v_{vir}$ is the virial velocity of the galaxy cluster halo,  as a function of the cluster-centric distance, provides  information on  whether galaxies are falling in (negative value) or moving outwards (positive value)  their host clusters.  

Figure~\ref{fig2} shows the number fraction of galaxies  (green histogram) in our sample as a function  of  the ratio $v_r/v_{vir}$. In order to highlight the radial distance dependence, we split the galaxy sample in four blocks depending on the cluster-centric position of the galaxies. The two first blocks focus on the inner regions of the cluster, thus,  a first block contains all galaxies with distances to the centre of their host cluster within 0.25 times the virial cluster radius, and the second block contains galaxies within $0.25 < r/r_{vir} < 0.5$. The outer part of the cluster is described by a third block with galaxies located at radial distances satisfying the condition $0.5 < r/r_{vir} < 1.0$. Finally, a fourth block describes the behaviour of galaxies close to a cluster but beyond the virial radius. 
In addition,  the blue (red) histogram in Fig.~\ref{fig2} stands for the fraction of gas-rich (-poor)  galaxies as a function of the ratio $v_r/v_{vir}$ for the four cluster-centric distance ranges.

\begin{figure}
\includegraphics[scale=.4]{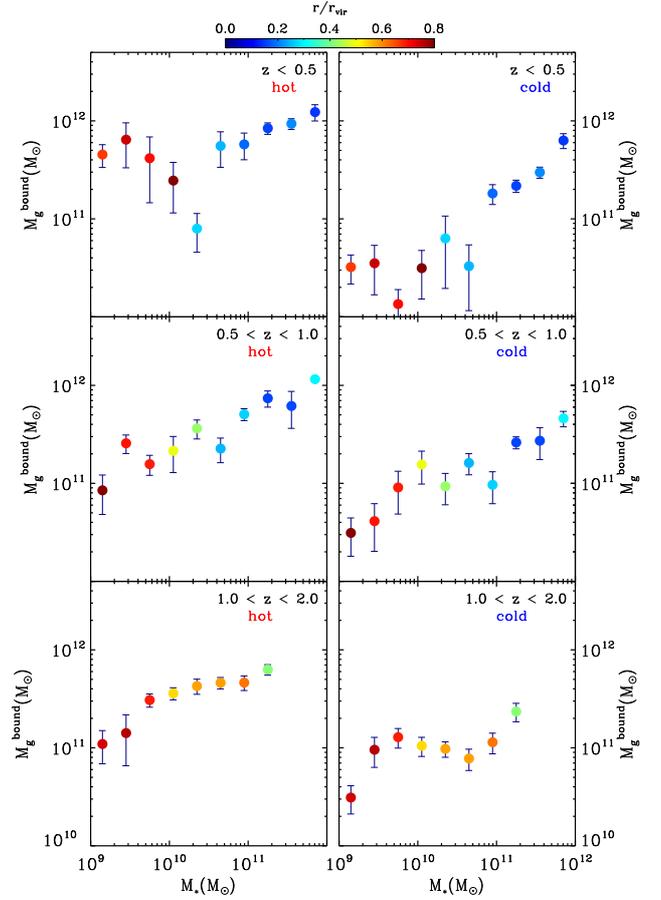}
\caption{Redshift evolution of the mass of bound gas ($M_g^{bound}$), hot (T>$5\times10^4$ K; left column) or cold (T<$5\times10^4$ K; right column), as a function of galactic stellar mass, $M_*$. Points and colours have the same meaning than in  Fig.~\ref{fig4}. }
\label{fig7}
\end{figure}

\begin{figure}
\includegraphics[scale=.4]{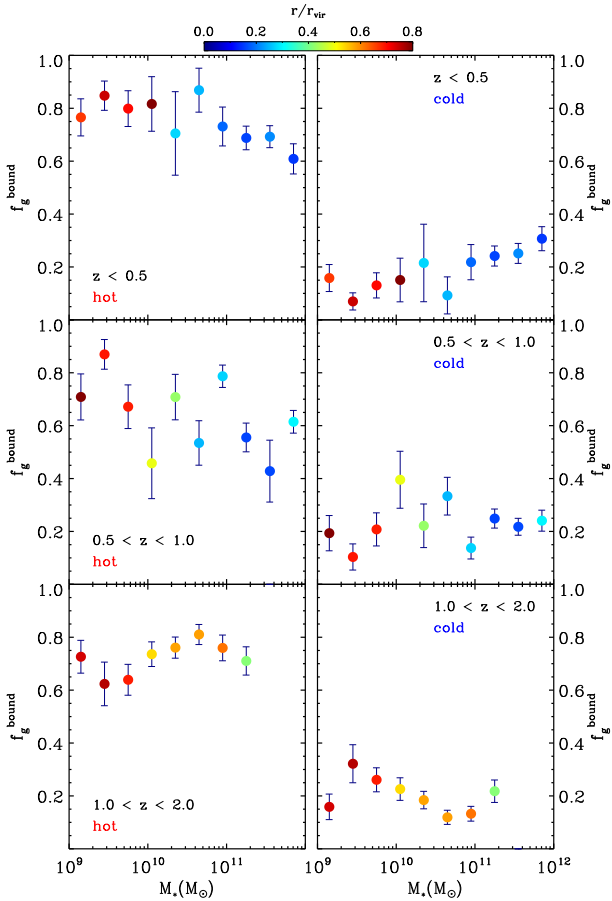}
\caption{Same as Fig.~\ref{fig6} but for the fraction of bound gas, $\rm f_{g}^{bound}$. }
\label{fig8}
\end{figure}

The analysis of Fig.~\ref{fig2} shows several interesting qualitative trends. In the inner part of clusters (top-left panel), the total number of galaxies moving inwards and outwards is roughly the same, and most of the galaxies have small radial velocities compared with the virial velocity of their host cluster. Gas-rich galaxies, being the dominant population of the sample, behave following the same line. There is also a small fraction of  gas-poor galaxies that follows a similar pattern. Although with a very poor statistical significance, there is a slightly larger number of gas-poor galaxies with positive ratios $v_r/v_{vir}$, that might indicate that theses objects have already travelled through the centre of the cluster. 
The top-right panel, representing galaxies with cluster-centric distances between 0.25 and 0.5 times the virial radius of their host cluster, shows a different scenario: galaxies in this radial regime  have a more widely spread range of radial velocities,  with a clear peak at small negative values ($v_r/v_{vir}\sim -0.1$). This result points  out the existence of an important number of galaxies falling radially towards the cluster centre. In this radial range, the whole subsample of  gas-poor galaxies shows also a maximum  at  $v_r/v_{vir}\sim -0.1$.
Within the outer half virial radius (bottom-left panel), gas-rich galaxies  are mainly following infalling radial trajectories with their velocities evenly distributed from negative to small positive values. Although not shown in the figure because they are out of the plot range, in  this radial region there is a small fraction of gas-poor galaxies, with high radial velocities, that corresponds to  a population of galaxies that have moved out from the cluster centre and that have lost their gas due to the interaction with the cluster central environment.  
Finally, the bottom-right panel of Fig.~\ref{fig2} shows the results for the galaxies located at the external neighbourhood of clusters. Like galaxies at the outer parts of clusters, most of the objects in this region are mainly infalling into the cluster potential with negative velocities.

The information on the dynamics of the galaxies obtained from the analysis of their radial velocity component can be complemented by looking at their tangential velocities ($v_t$). In particular, in Fig.~\ref{fig3}, we present the same analysis than in Fig.~\ref{fig2} but as a function  of the ratio of the absolute values of the tangential to radial velocities. In this figure, low values of the ratio $|v_t|/|v_r|$ stand for strong radial motions, whereas high values represent eccentric trajectories.  

Most of the gas-rich galaxies located at inner cluster regions (top-left panel) show a clear tendency towards dominated radial motions, with a certain dispersion that becomes more significant for larger cluster-centric distances. 
In the range of distances from 0.25 to 0.5 times the virial radius of the cluster (top-right panel), the sample shows the largest amount of galaxies with eccentric orbits. Galaxies in the outer half virial radius (bottom-left panel) present a similar behaviour to the ones in the previous radial regime, showing a large variety of tangential velocities. Finally,  gas-rich galaxies at the outskirts of clusters (bottom-right panel) present a slight tendency to radial trajectories.   
Whereas gas-rich galaxies  exhibit a wide range of values of the ratio of tangential to radial velocities, gas-poor galaxies can be grouped in two broad populations, one prone to radial velocities, and a second one with values of the velocity ratios close to one. This trend happens at all radial ranges except in the outer half virial radius, where there are several objects with eccentric trajectories.  Despite the low number statistics, it is significant that our sample of gas-poor galaxies follows in all cases a similar and general trend.

\subsection{Interaction with the environment}

As a result of the interaction with the environment, ram-pressure stripping can play a crucial role in removing the gas component in galaxies. In this section, we present the results of the analysis of our sample of galaxies in order to assess the effects of the ram-pressure stripping. 

As introduced in Sec.~\ref{method}, we define the amount of gas susceptible of being removed by RPS as the total amount of gas in a galaxy and its halo  that is not gravitationally bound. To fulfill this condition, we look at the gas  within each computational cell as an elementary  parcel of gas. Whether the gas within a computational cell is gravitationally bound or unbound is decided by computing the total energy, $E_t$, of this parcel of gas, namely, the kinetic plus gravitational energies. The parcels of gas with $E_t\ge 0$ are labelled as unbound gas. Therefore, all cells satisfying the previous condition are volume elements of gas that, although within the galactic halo, are in process of being removed.

\subsubsection{The unbound component: gas removed from galaxies}

With the previous definition of removable gas, it is possible to compute the changes in this component as a function of time.   
Figure~\ref{fig4} displays the average mass of unbound gas, $M_g^{unbound}$ (right column), and the mass fraction of unbound gas, ${\rm f_g^{unbound}}$ (left column),  for all galaxies binned as a function of the stellar mass. The three rows of panels in Fig.~\ref{fig4} show the results for three different redshift ranges.  The error bars represent the standard deviations around the mean value for the galaxy subsample within each stellar mass bin. At the same time, 
symbols are colour-coded according to the mean cluster-centric distance in each mass bin. 

The analysis of Fig.~\ref{fig4} allows us to discuss the evolution of galaxies depending on their stellar masses and positions within their host clusters. Visual inspection of this figure provides a clear distinction between galaxies with stellar masses above and below $10^{10}\, M_{\odot}$. 
At early stages of evolution ($1.0< z < 2.0$),  low(high)-mass galaxies, $M_* \lsim 10^{10}\, M_{\odot}$ ($M_* \gsim 10^{10}\, M_{\odot}$),  
show similar dependence of the unbound mass on the galaxy stellar mass, but with different normalization. For both groups of galaxies, the mass of unbound gas can be linearly fitted by  
$log(M_{g}^{unbound})=(0.63\pm0.12)log(M_*) + (4.20\pm1.25)$ for lower masses, and 
by  $log(M_{g}^{unbound})=(0.68\pm0.09)log(M_*) + (3.29\pm0.93)$ for more massive systems.
Given the low statistical significance of our galaxy sample, we cannot conclude that both galaxy groups show different behaviour since their fittings are compatible with each other within the error bars. Nevertheless, these results seem to suggest possible differences in the amount of unbound gas in galaxies with stellar masses larger or smaller than $10^{10}\, M_{\odot}$.
We interpret this different trend as a consequence of the swallower gravitational potential well of the smaller objects, which would retain less gas, specially,  gas within the galactic halo.

At redshifts between 1.0 and 0.5, the characteristic stellar mass of $10^{10}\, M_{\odot}$ clearly marks two different behaviours of the galaxies in the sample.  
Thus, the low-mass galaxies have barely changed their amount of unbound gas, maintaining very similar values to previous epochs. 
At the same time, their location within the cluster remains almost unchanged. 
Above $10^{10}\, M_{\odot}$, all systems have moved towards more central locations within the cluster and, in this process,  they have increased the amount of unbound gas in their haloes. This tendency is stronger in the most massive systems.

The top-right panel of Fig.~\ref{fig4} shows the results for the lowest redshift range, i.e, $z < 0.5$. At this epoch, the low mass galaxies present very similar amount of unbound gas regardless of their stellar mass. 
There is a visible break in the trends in terms of position and amount of unbound gas marked by the characteristic  stellar  mass of  $10^{10}\, M_{\odot}$.
On average, low-mass galaxies exhibit a tiny drop in the mean amount of unbound gas. At the same time, the radial locations of  these galaxies within the cluster are roughly the same than at previous epochs. On the contrary, the group of massive galaxies keeps on getting closer to the centre of the cluster. The amount of unbound gas of this group can be fitted by 
 $log(M_{g}^{unbound})=(0.65\pm0.09)logM_* + (3.38\pm1.01)$. This correlation is very similar to the one existing at  earlier times (see bottom-right panel of Fig.~\ref{fig4}).

 The left panels in Fig.~\ref{fig4} show the fraction of unbound gas, $\rm f_{g}^{unbound}$, defined with respect to the total amount of gas
within the galactic haloes for galaxies at the same redshift intervals. This information  is complementary to the amount of unbound gas shown in the right panels of the same figure. At early epochs, 
the mean value of the gas fraction of unbound gas, $\rm f_{g}^{unbound}$, for galaxies with $M_* < 10^{10}\, M_{\odot}$ is $0.09\pm0.03$, and 
$0.07\pm0.03$ for the larger systems. Thus, the two groups of galaxies show fractions of unbound gas that are compatible within the errors.

At intermediate redshifts (middle-left panel), there are not clear trends and the gas fraction of unbound gas is roughly constant for all galaxies ($\sim0.14\pm0.09$) independently on their stellar mass. At low redshifts (top-left panel), 
the scatter of previous epochs gets reduced, and the two groups of galaxies seem to show a slightly different behaviour, with the low-mass galaxies presenting a fraction of unbound gas almost constant, and the massive systems pointing towards a weak correlation with the stellar mass of the galaxies. The mean value of $\rm f_{g}^{unbound}$ is  $0.06\pm0.02$.

 We show in Fig.~\ref{fig5}, 
the average radial position of the galaxies -- in units of the virial radius of the host cluster --  as a function of their stellar mass
for the same three epochs displayed in Fig.~\ref{fig4}. The error bars represent one standard deviation. The numbers below each bin stand for the number of galaxies contained within a given stellar mass bin. 
As a general trend, at early times, all galaxies are quite evenly distributed throughout the cluster, being, however, the less massive systems mainly located at the most external cluster regions. As time evolves, galaxies with masses larger than $10^{10}\, M_{\odot}$ move towards more central positions within the clusters being this trend more notorious for the most massive systems. At low redshifts, 
differences according to the stellar mass of the galaxies are already evident.  Thus,  galaxies with  $M_* < 10^{10}\, M_{\odot}$ are located at radial distances larger than half virial radius, whereas galaxies with $M_* > 10^{10}\, M_{\odot}$ occupy  positions within the cluster centres closer than 0.4 virial radius. 

The correlation between the stellar mass of more massive galaxies and their amount of unbound gas does not seem to depend strongly on the position of those galaxies. Results from Fig.~\ref{fig5} show how, even when the position of massive galaxies can vary substantially, the correlation between total stellar mass and unbound gas mass is found. 
We interpret this result as a consequence of the correlation between the stellar mass and the total gas mass.
This translates into a rather constant trend in the value of $\rm f_{g}^{unbound}$.

So far, we have not studied the differential effect of the environmental interaction on the gas at different thermodynamical states.
In order to study this scenario, we split the gas in galaxies in two phases: the cold gas component, including all cells with gas temperature  $T<5\times 10^4\, K$, and the hot gas component, that stands for all cells with temperature $T>5\times 10^4\, K$.
According to this criterion, we compute for all galaxies in the sample the mass of hot and cold unbound gas.  With these quantities, we introduce two new relative fractions, $\it f_{g}^{unbound}$,  defined as the mass of hot (cold) unbound gas over the 
total mass of hot (cold) gas within the galactic halo. The main difference with respect to the previously introduced fraction, $\rm f_g$ (defined with respect to the total amount of gas), is that the relative fraction, $\it f_g$, stresses
the diverse evolution of each component (cold and hot), highlighting whether one of the phases is  more efficiently stripped.

Figure~\ref{fig6} shows the results for the relative fractions of hot and cold unbound gas, $\it f_g$. The hot (cold) component is displayed in the left (right) column. The redshift ranges and the colour dots have the same meaning than in Fig.~\ref{fig4}.
The analysis of Fig.~\ref{fig6} reveals that cold gas is more affected by the interaction with the environment. This effect is even larger for massive galaxies, where the relative fraction of stripped cold gas can reach values as high as $40\%$. This could have important implications in reducing the star formation.

As a summary, we find that gas removal is not very effective, being the fraction of unbound gas always below $30\%$. We do not notice any evidence for a significant dependence of the fraction of removable gas on the stellar mass. On the contrary, the unbound gas mass 
strongly correlates with the stellar mass. The fraction of removable gas is higher at intermediate redshift ($0.5<z<1$), 
although the poor statistics prevents us to draw a robust conclusion on this respect. Finally, the analysis of the relative fractions reveals that the interaction with the environment is more dramatic on the cold gas and,  specially, in massive systems, where the fraction of cold unbound gas can be soaring.

\subsubsection{The bound component: the gas retained by galaxies}

Following a similar approach to the study of the unbound gas component, we focus now on the gas that  is kept in galaxies, namely, the bound component. 
The amount of gas maintained by galaxies is a key quantity that determines their evolution and has direct effects on many physical processes like the star formation. However, the role of this gas would be different depending on its temperature. Thus, 
in order to study this component,  we adopt the same approach than in previous Section and split the bound gas into two phases: cold ($T < 5\times10^4\, K$) and hot ($T > 5\times10^4\, K$). 

The masses of hot and cold bound gas have been computed for all galaxies in the sample. The results are displayed in Fig.~\ref{fig7}.   
The left (right) column shows the hot (cold) mass of bound gas for all the galaxies at three different redshift intervals as a function of the galactic stellar mass. The dots in the plot stand for the average value of the galaxies in each stellar mass bin, and the error bars represent one standard deviation. At the same time, the dots are coloured according to the mean cluster-centric distance of the galaxies in each stellar mass bin (see Fig.~\ref{fig5}).

The analysis of Fig.~\ref{fig7} leads to the following results. As it has been already described in Fig.~\ref{fig5}, at early stages of evolution all galaxies are distributed evenly throughout the cluster with no clear preference of location. The characteristic mass of $M_*\sim10^{10}\,M_{\odot}$ is again  separating the behaviour of galaxies. Thus, the low mass galaxies present an increasing amount of bound, both hot and cold, gas as a function of their stellar mass. However, systems with stellar masses above to $10^{10}\,M_{\odot}$ present almost constant values of the amount of bound gas, regardless of the stellar mass of their host galaxies. On average, the amount of hot bound gas is $\sim 4$ times larger than the mass of cold bound gas. 

\begin{figure}
\centerline{\includegraphics[scale=.3]{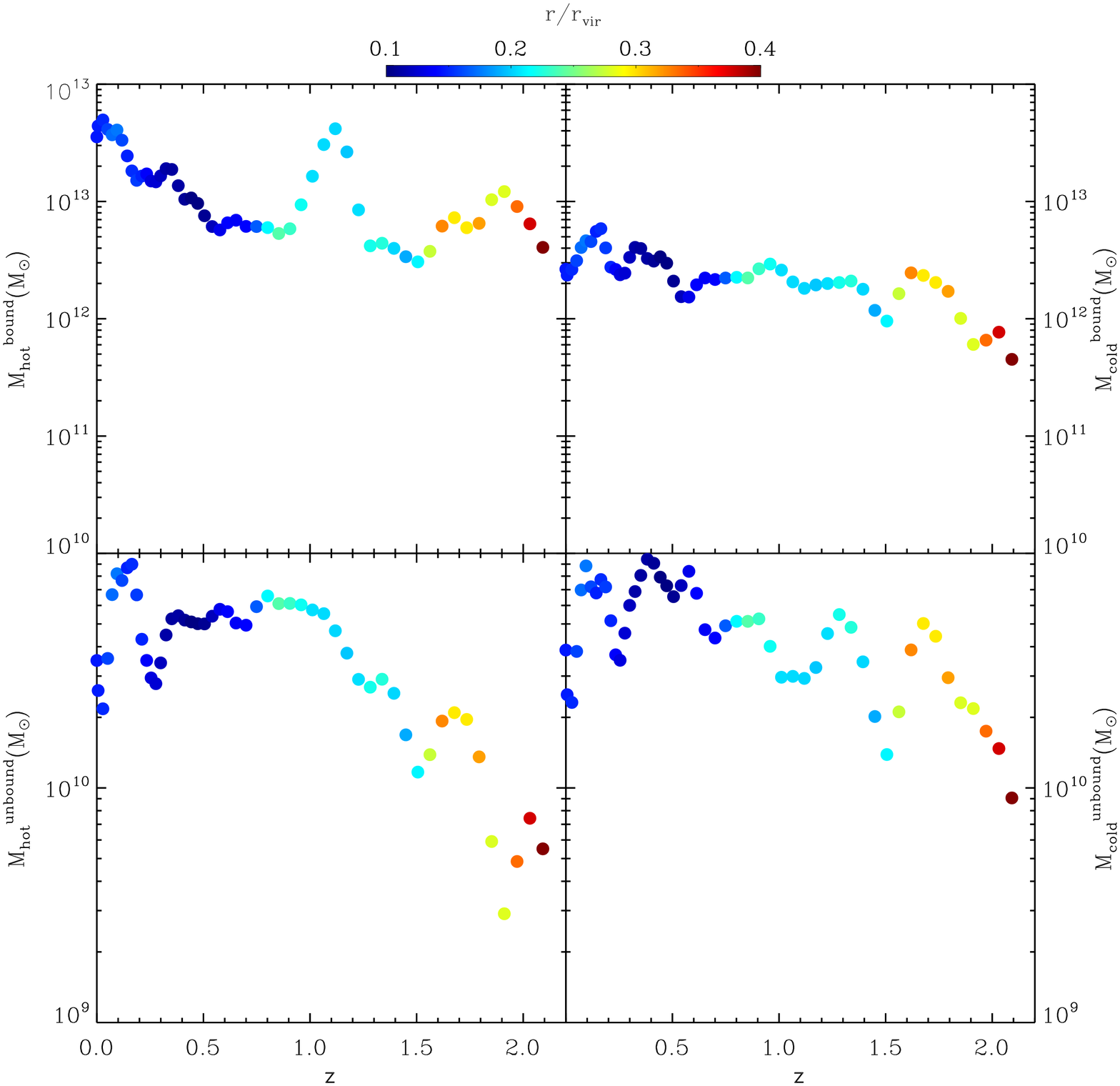}}
\caption{ Evolution of the gaseous mass content (bound and unbound) averaged for 
 all  galaxies whose history can be tracked back from $z\sim 0$ to $z\sim 2$. The masses of the four gas components,  namely, 
 hot (T>$5\times10^4$ K) bound gas (top-left panel), cold (T<$5\times10^4$ K) bound gas (top-right), hot unbound gas (bottom-left), and cold unbound gas (bottom-right) are displayed. At a given redshift, points are coloured according to the average radial distance of the galaxies (in units of the cluster virial radius) to centre of their host galaxy cluster.}
\label{fig9}
\end{figure}

During the evolution, the behaviour of both gas components, namely hot and cold, becomes distinguishable. Thus, the right middle panel in Fig.~\ref{fig7} shows a mild increase with respect to previous times of the amount of cold bound gas in the galaxies with intermediate stellar masses.  The values of this quantity are roughly constant although with a larger scatter than at previous epochs. On the contrary, the low-mass galaxies do not show any relevant change in this quantity. As for the unbound component at this epoch (see Fig.~\ref{fig4}),  the high-mass galaxies emerge showing the highest values for the mass of cold bound gas. 

The hot bound gas at $z$ between 1.0 and 0.5 presents a different behaviour. The differences begin for galaxies with stellar masses above $10^{10}\,M_{\odot}$. Thus, galaxies with intermediate stellar masses have quantities of hot bound gas that seem to increase with the stellar mass of the galaxy. This effect is not so clear for the cold gas component. Moreover, whereas the cold component slightly increases, the amount of hot bound gas gets slightly reduced in this redshift interval compared to the previous one. For the most massive galaxies, there is a clear correlation. Thus, the larger the stellar mass of the galaxy, the greater the amount of hot bound gas. 

The more dramatic changes appear at low redshifts.  In this regard, for the cold gas component, the changes mainly affect  the intermediate mass galaxies that suffer an important reduction of their amount of cold bound gas compared with previous epochs, whereas the high and low mass galaxies barely change.  In the hot component, there are noticeable increases of the amount of hot bound gas in both, low and high mass galaxies. At  intermediate mass ranges, the plot shows a remarkable dip in the amount of hot bound gas  for the galaxies 
with stellar masses slightly above $10^{10}\,M_{\odot}$.

One noticeable feature in Fig.~\ref{fig7}  is the ratio between the amounts of hot and cold gas components. Thus, almost at all redshifts and for all stellar masses, the quantity of hot bound gas is between a factor of $\sim 3-4$ larger than the amount of cold bound gas. However, this trend is broken at low redshifts, specially for the galaxies with 
$M_* < 10^{10}\,M_{\odot}$, which can have a factor of  $\sim 10$ times more hot  than cold bound gas.
  
As for the unbound gas, we also analyse the fractions of hot and cold bound gas, defined as the ratio of mass of bound gas to the total gaseous mass in the halo. This result is presented in Fig.~\ref{fig8}, which is similar to Fig.~\ref{fig7}. 
The fraction of bound gas, both hot and cold, at early times is rather constant independently of the stellar mass of the galaxies. This behaviour is similar to the one obtained for the total amount of unbound gas mass at the same epoch (see lower panel in Fig.~\ref{fig7}).  Moreover, the fraction of hot bound gas is roughly a factor of four larger than the fraction of cold bound gas.

In the middle panels of Fig.~\ref{fig8}, at redshifts between 1 and 0.5, the fractions of bound gas present an opposite behaviour than the one obtained for  the total amount of bound gas. Meanwhile the total amount of hot and cold bound gas increases with the stellar mass, the fraction of bound gas is compatible with a constant value regardless the stellar mass of the galaxies or, in the case of the hot component, with a small decrease as a function of $M_*$. Compared with the previous period of time, the fraction of cold bound gas has not barely changed, whereas the fraction of hot bound gas is slightly smaller. 

At low redshifts, there is a weak tendency of the fraction of cold bound gas to increase with the mass of the galaxy, whereas the fraction of hot bound gas seems to be constant for all galaxies except for the larger systems, where there is a lowering of the values of $\rm f_{g}^{bound}$. The average values of the fractions are very similar to those at early epochs. Indeed, the average values of the fraction of hot bound gas is roughly four times larger than the fraction of cold gas. In the case of the most massive systems, this difference gets reduced to just a factor of two.

\subsection{Individual galaxy histories}

In previous sections, we have taken a statistical approach by considering the average properties of all our sample of  galaxies at different epochs. In this Section, we follow the time evolution of individual galaxies and focus on the fate of their gas.

The procedure is as follows:  first, all  galaxies in our sample at $z\sim 0$ are identified; then,  the history of each individual galaxy is reconstructed tracking it back until it is first identified by the halo finder; finally,  once each galaxy is identified individually at all redshifts, all relevant quantities are computed as a function of $z$. 

As in previous sections, we classify the gas content,  within two half-mass radius of each galaxy,  into 
four different components: i) hot ($T > 5\times10^4\,K$) bound gas, ii) hot  unbound gas, iii) cold ($T < 5\times10^4\,K$) bound gas, and iv) cold  unbound gas. Similarly to the previous definitions, a parcel of gas is considered to be unbound when its total energy is equal or larger than zero. 

We have identified eight galaxies at $z\sim 0$ that can be tracked backwards in time, but only one was already formed at $z\sim2$. 
In Figure~\ref{fig9}, we present the redshift evolution, since $z\sim2$ down to $z=0$, for the mass  of hot bound gas (top-left panel), cold bound gas (top-right), hot unbound (bottom-left), and cold unbound gas (bottom-right). 
In Fig.~\ref{fig9}, the circles represent the average value of the mass of each gas component for all galaxies at each redshift interval.  As in previous figures, in addition to the time evolution of the different gas components, the dots are coloured according to the average cluster-centric distance of the galaxies (in units of the cluster virial radius). 

Figure \ref{fig9} shows different behaviours depending on whether we deal with the hot or cold gas component. 
At early times, most of the gas in galaxies, whose histories can be reconstructed backwards from $z\sim 0$ to $z\sim 2$, is formed by bound gas, mostly hot. The unbound component is subdominant, being the amount of cold unbound gas the less abundant of all the four components. On average, all the galaxies are initially located at an average distance of around $0.4$ virial radius. 
As time goes on and  galaxies get closer to the centre of their host clusters, the four gas components evolve in different manners. Basically, the amount of all  four components increases with time but with a quite different pace. 
Indeed, as evolution proceeds,  galaxies move within their host galaxy clusters but, at the same time, they also evolve increasing, therefore, their mass and accreting hot gas from the ICM. A fraction of the accreted hot gas is mixed up with  the hot gas in the galactic halo, whereas a fraction of this gas eventually cools and becomes part of the cold component. At the same time than these processes are taking part, galaxies interact with their environments in the clusters. This mechanism is able to remove gas, hot and cold, from the galactic halo. 

The evolution of the cold gas is very similar for the bound and unbound components. From $z\sim 2$, the amount of cold gas has an increasing trend modulated by several oscillations. The bound component is always larger than the unbound one by a factor $\sim 4$.
This proportionality suggests that, on the one hand, the cold gas is replenish at a smooth rate related with the cooling of the hot gas. On the other hand, at all epochs, around 25\% of this cold gas can be removed from the galaxy. The maximum removal of cold gas is produced at $z<0.5$, coinciding with the time when the galaxies get closer to the cluster centres. 

The hot gas component behaves in a different manner. Thus, the hot bound gas does not evolve so smoothly as the other components, and presents strong oscillations with several local maxima at $z\sim 1.9$ and $z\sim 1.1$. Moreover, the mass of hot unbound gas increases quite smoothly, compared with the bound component, up to $z\sim 0.7$ when remains fairly constant until lower redshifts. The ratio of hot bound gas to hot unbound gas varies from values of around 4 out to more than 100 at some epochs. 

\begin{figure}
\centerline{\includegraphics[scale=.45]{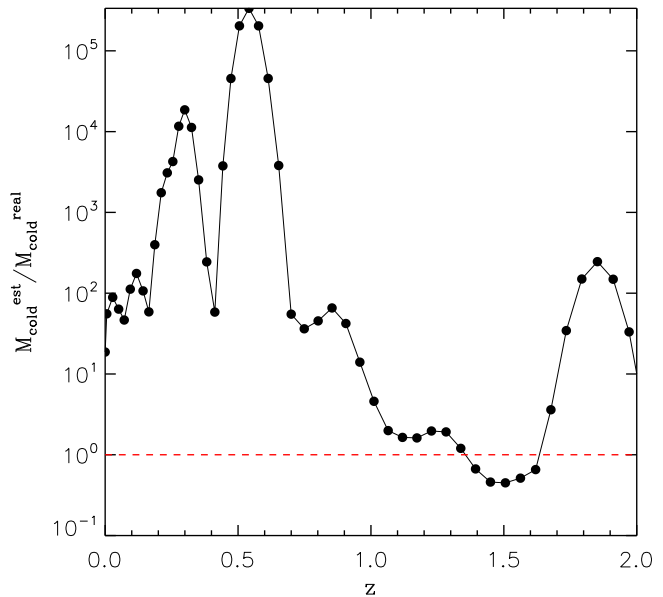}}
\caption{Average ratio of the estimated mass of cold bound gas to the real mass of cold gas as a function of redshift for all galaxies whose histories can be reconstructed backwards in time from  $z\sim 0$ out to $z\sim 2$. }
\label{fig10}
\end{figure}

The evolution of the cold gas component is more complex to describe as it is affected by star formation, that depletes part of this
component. A certain amount of cold bound gas must be locked into stars formed along the galaxy evolution and, therefore, is not included in previous results. This would imply a reduction of the fraction of cold bound gas that has to be disentangled from the similar effect produced by RPS. 
In order to asses  the role of the ram-pressure stripping on the cold gas component and the star formation consumption of this component, we present Fig.~\ref{fig10}. In this plot, we estimate, for a given $z$, the amount of cold bound gas that should exist, $M_{cold}^{est}$, if the effects of RPS were not relevant. To do this, we start by considering the amount of cold bound gas at a previous redshift, $M_{cold}(z-\Delta z)$. 
If there were no RPS, this quantity should increase due to cooling of part of the hot bound gas. This increment can be estimated as 
the mass of hot bound gas $M_{hot}$ at $z-\Delta z$ minus the mass of hot bound gas $M_{hot}$ at $z$, where we have assumed that, for a small enough $\Delta z$, the amount of hot gas accreted from the ICM that has turned into hot bound gas can be neglected. To this quantity, that provides the predicted amount of hot bound gas moved into cold bound gas, it has to be subtracted 
 the amount of hot unbound gas and the increase of the stellar mass at the considered redshift interval $\Delta z$.	
 The $M_{cold}^{est}$ can be computed as follows:
\begin{eqnarray}
M_{cold}^{est}(z)&=& M_{cold}^{bound}(z-\Delta z) \nonumber \\
&+& [M_{hot}^{bound}(z-\Delta z)- M_{hot}^{bound}(z)] - M_{hot}^{unbound}(z)\nonumber \\
&-& [M^*(z)-M^*(z-\Delta z)].
\label{mc_esti}
\end{eqnarray}
\noindent
The predicted change in the cold bound component for each galaxy, $M_{cold}^{est}$, can be compared with the real change in this component, $M_{cold}^{real}$, obtained by comparing its value at $z$ and at $z-\Delta z$. In an ideal case with no RPS effect at work, the  quantity $\frac{M_{cold}^{est}}{M_{cold}^{real}}$ should be equal to one or close to one, as  some assumptions have been made when computing it. Values smaller than one would indicate very active episodes of star formation or that the approximations made to compute $M_{cold}^{est}$ are not well suited. On the contrary, large values of $\frac{M_{cold}^{est}}{M_{cold}^{real}}$ indicate that the amount of cold bound gas left is smaller than expected if only cooling and star formation were acting, that is, the presence of RPS. The absolute 
values of this ratio must be taken with caution, as this is only an approximate method to estimate which of both processes controls the removal of cold gas.  

This effect is studied in Fig.~\ref{fig10}, where the average value of $\frac{M_{cold}^{est}}{M_{cold}^{real}}$ for all galaxies in our sample is shown as a function of $z$.  This plot shows that, in our simulation, although star formation must play a role consuming cold bound gas, the RPS is the dominant mechanism removing this gas component during the whole time evolution of the galaxies in our sample, as the real amount of cold gas is much lower than expected by simplified calculations. 

Our simulation does not include AGN feedback and, therefore, it can be discussed whether the star formation rates in our sample of galaxies are realistic \citep[e.g.][]{Planelles_2013}. We do not expect that these limitations could have a relevant impact on our results on the RPS effects, though, as it is shown in Fig.~\ref{fig10},  the changes produced by the star formation activity should be dramatic in order to substantially affect the outcome of the interaction between galaxies and their environment. 
Thus, the trend shown in Fig.~\ref{fig10} could be considered as a conservative estimate of the role of the RPS.

\begin{figure*}
\centerline{\includegraphics[scale=0.9]{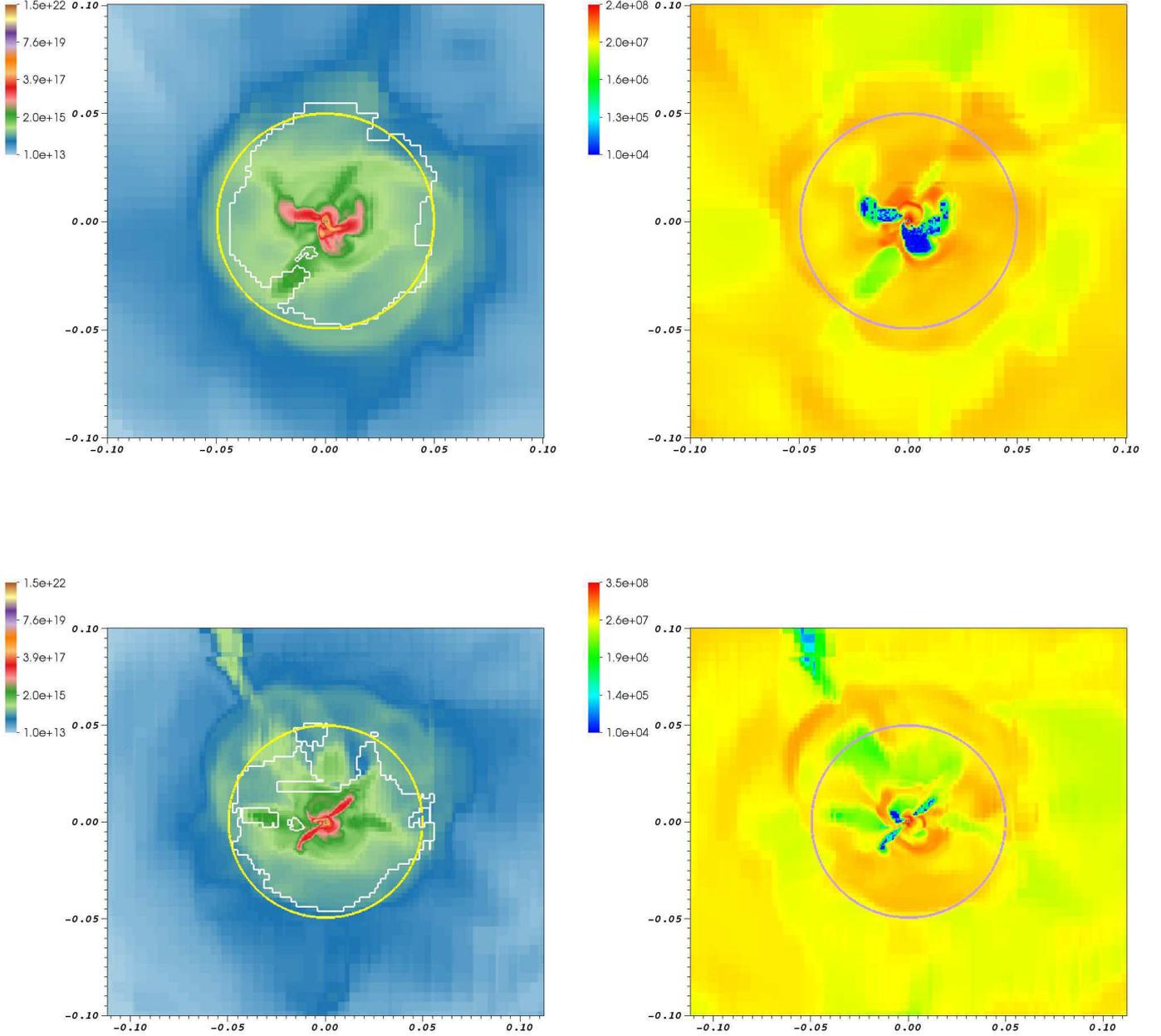}}
\vspace{-2.8cm}
\caption{Density (left panels) and temperature (right panels) of the gas in a thin slice through the centre of one of the best numerically resolved galaxies in the simulation at $z\sim 0$. The upper (lower) panels represent a face-on (edge-on) view of the galaxy. The density (temperature) of the gas is in units of $M_{\odot}/Mpc^3$ (K). The yellow and purple circles mark the radius of the galactic halo. The white contour defines the boundary between bound and unbound gas.The axis of the panels are in units of $Mpc$.}
\label{fig11}
\end{figure*}

\subsection{Detailed description of the effects of the interaction with the environment}

\begin{figure*}
\centerline{\includegraphics[scale=.95]{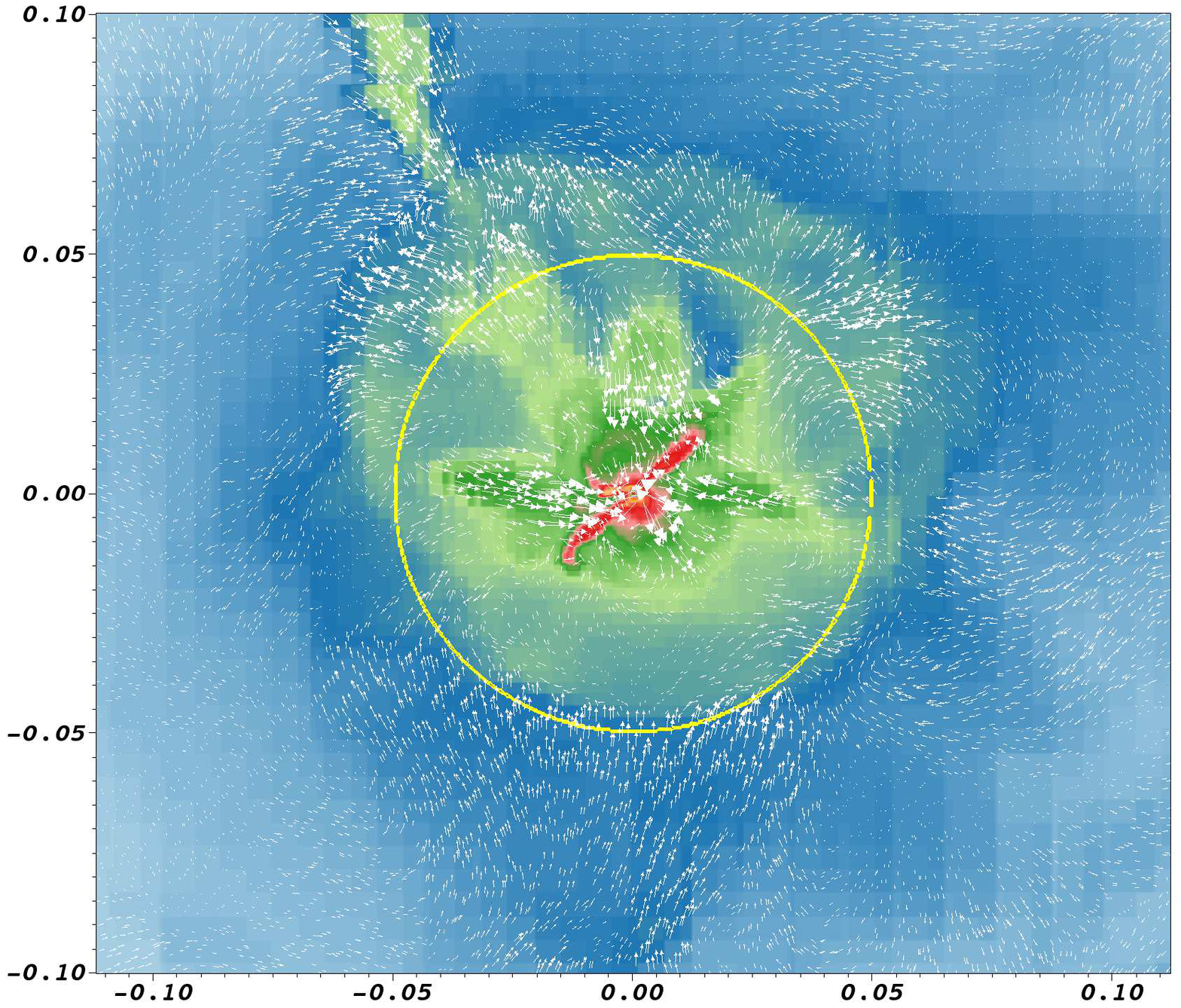}}
\vspace{-4.0cm}
\caption{Gas density in a thin slice through the centre of a galaxy from a edge-on view. Superimposed to the density map, the velocity field of the gas is shown as white vectors whose size is scaled according to their magnitudes. The galaxy is moving towards the right lower corner of the image. The axis of the panels are in units of $Mpc$. The velocity frame of reference is the centre of the computational box}
\label{fig12}
\end{figure*}

In order to illustrate the processes described in previous Sections, we present a detailed study of a prototypical galaxy in our simulation. 
The chosen system is located close to the centre of its host galaxy cluster at $z\sim 0$, when  this galaxy is one the best numerically resolved in the simulation.  Its stellar mass is $M_* \simeq 1.7\times 10^{12} \, M_{\odot}$, and its half-mass radius is   $R_{1/2}\simeq 24.8 \, kpc$. This galaxy is in a galaxy cluster with total mass  
$M_{vir}\simeq 1.1\times 10^{14} \, M_{\odot}$, virial radius $r_{vir}\simeq 1.25 \, Mpc$, and virial velocity $v_{vir}\simeq 350.5 \, km/s$. The galaxy is located at $175\, kpc$ $(0.14r_{vir})$ from the cluster centre. Its relative velocity with respect to the cluster centre of mass is $\sim 90\, km/s$. The radial component of its velocity is negative, indicating that the galaxy is infalling into the cluster. However, its trajectory is  far from being radial as the ratio of the absolute values of the tangential and radial velocities is $\sim 1.1$.

Figure~\ref{fig11} displays four thin slices of two different views of this object. The left (right) panels show the gas density (temperature), whereas the upper panels stand for a face-on view of the galaxy and the lower panels present the edge-on of the galaxy. In all cases, the radius of the galactic halo is represented as a circle. For the sake of clarity, this circle is yellow (purple) on density (temperature)  maps.
The white contour represents the boundaries between bound and unbound gas. 

Thus, the face-on density map shows how most of the gas in the halo is bound with the exception of a few holes. Similar effects can be seen at the edge-on view (right lower panel).

This object is a disk-like galaxy with prominent spiral features. 
In the face-on view,  the gas density maps clearly shows spiral features, whereas the right lower panel shows a very thin and dense disk at the centre of the halo. This spiral features in the density maps are also noticeable in the temperature plots (left panels). The colder regions are clearly correlated with the thin gaseous disk. 
 
 The lower panels in Fig.~\ref{fig11} reflect the features of the movement of this galaxy with respect to its environment. The direction of movement of this galaxy, using as reference frame the panels, is towards the right lower corner. This movement translates into a gas compression ahead the direction of movement that it is also seen as a gas heating in the temperature map. Behind the galaxy, a wake is left with regions of cooler and low density gas. Even at regions further away behind the galaxy centre, a remarkable shell of dense and hot gas shows how the gaseous galactic halo has been pushed away by its interaction with the environment. 
 
One of the most relevant features of these plots is that this interaction with the environment produces a quite inhomogeneous picture, with hot and underdense regions mixed with colder and denser blobs of gas. This issue points towards the fact that the interaction of a galaxy with its environment is an extremely complex scenario which can not be described by a simplified model.

To deepen in this last point, we present Fig.~\ref{fig12}. This figure shows the density map for a edge-on view of the galaxy that corresponds to left lower panel of Fig.~\ref{fig11} where the velocity field of the gas in this slice has been superimposed. The velocities, shown as white vectors, confirm the details of the interaction between the galaxy and its environment. 

The ICM gas ahead the galaxy shows very inhomogeneous pattern, alternating regions with strong flows almost orthogonal to the disk plane of the galaxy, close to regions with very low velocity gas or even areas with vortexes. Within the galactic halo, the situation is even more complex. Thus the outer regions of the halo ahead the galaxy present most of the gas with very low velocities. As we get close to the disk, a complex pattern of flows  moving in and out arise. A similar structure can be seen behind the disk, where strong ingoing flows coexist  with almost stopped gas and outgoing flows. It is remarkable the shell of gas belonging to the galactic halo pushed away by the interaction with the environment. 

\section{Discussion}

In this paper, we study the effect of the interaction with the environment for a sample of galaxies obtained  in a cosmological simulation carried out with an AMR code. In order to asses the consequences on the gaseous content of galaxies in clusters when they interact with their environment within the galaxy clusters, we propose a method based on estimating the total energy of all elementary gas volumes. This method has several advantages. It is local and it can be applied to any numerical element in which the simulated galaxy is resolved. It can be applied to non idealized galaxies obtained in a fully cosmological simulation without the need to assume any particular geometry  or shape of the volume wrapping the studied galaxy. Besides, it is a well motivated physical condition to know whether a parcel of gas in a physical structure is bound or unbound to the system where they are located.  

The galaxies in clusters in the final stages of our simulation have a tendency to be located towards the centre of the galaxy clusters at low redshifts. Those ones in central parts of the clusters show mostly radial orbits with low velocities compared with the virial velocity of their host galaxy cluster. This tendency is even more marked for gas poor galaxies, indicating that these systems have been processed by the clusters. On the contrary, the galaxies located at the outer parts of the clusters exhibit more eccentric orbits.  Some works \citep[e.g][]{Cen2014} suggest that an important fraction of galaxies can be affected by their interaction with the environment at distances from the centre as far as 2-3 virial radii, pointing towards an early effect of  RPS. However, we have not found in our results any evidence of such early environmental effects. 

The study of the time evolution of the gaseous content of galaxies reveals that at early stages galaxies are evenly distributed within clusters with a strong correlation between the amount of unbound gas and their stellar masses. As time goes on, the galaxies migrate towards central  locations in the clusters. During this process, all galaxies have unbound gas, but there are strong differences depending on the stellar mass of the galaxies. Thus, whereas the amount of unbound gas for the low mass systems ($M_* < 10^{10}M_{\odot}$) is almost constant, it increases as a function of the stellar mass of the galaxy for more massive systems.  This tendency is maintained during the whole time evolution. 
The thermodynamical state of the gas also plays a relevant role so as to understand the removal of gas by interaction with the environment. Thus, when the gas is separated in two phases: hot ($T>5\times10^4\, K$) and cold ($T<5\times10^4\, K$), the relative fractions, defined as the amount of unbound -- hot or cold -- gas divided by the total amount of hot or cold gas, respectively, show relevant differences. The cold component is more efficiently removed than the hot gas, reaching values as high as $40\%$. 

The gas retained by the galaxies (bound component) have a different behaviour compared with the gas susceptible to be removed (unbound component) and it strongly depends on its temperature. 
The evolution of the cold bound  gas ($T<5\times10^4\, K$) strongly depends on the stellar mass and cluster-centric distance of the galaxy. 
The most massive systems present high amounts of cold bound gas strongly correlated with the stellar mass of the galaxy with minimal variations during the evolution, indicating that the effect of the interaction with the environment is not crucial on this component for such massive systems. This results can be understood due to the stronger gravity of these systems together with the efficiency of cooling and their low relative velocities in the clusters.   The low mass galaxies ($M_* < 10^{10}M_{\odot}$) suffer the opposite effect. The lower gravity due to their mass, the less effective cooling and their locations at the outer parts of cluster with low densities of the environment, produces the lowest fractions of cold bound gas. The effect of the interaction with the environment is more noticeable in the systems with intermediate stellar masses ($10^{10}M_{\odot} <  M_* <  10^{11} M_{\odot}$) where the amount of cold bound gas changes as the galaxies moves towards more central positions within the clusters. 

The analysis of the evolution of the hot bound component ($T>5\times10^4\, K$) is not so clear as it is strongly affected by a continuous replenishment of hot gas from the ICM. As a general trend, the amount of hot bound gas is a factor 3 or 4 times larger than the cold component. This ratio can be as high as 100 times for low mass galaxies, where few gas is able to cool and most of bound gas is hot gas from the corona.  

In order to present a complementary approach to the statistical analysis of the properties of galaxies, we have studied their individual histories. The results of such procedure reinforces the conclusions from the statistical analysis and leaves a general description of the process experienced by galaxies evolving inside the clusters. As the galaxies get closer to the centre of clusters, they accrete hot gas from the ICM that it is mixed with the gas in the galactic halo. Part of this gas cools and it is moved to the cold component. Simultaneously, as the galaxies travel inwards in the clusters, they interact with the environment and this interaction removes gas in both phases, producing hot and cold unbound gas. In all galaxies whose individual histories can be reconstructed, the unbound components are subdominant compared with the bound component.  Typically, the amount of cold bound gas is 4 times larger than the quantity of unbound gas. In the case of the hot unbound component, there is no such a clear trend, as this component is highly contaminated by the ICM gas.

The detailed analysis of the interaction of the galaxies with their environment reveals that this process is a rather complex scenario. The environments have turned out to be quite inhomogeneous mediums, with a complex pattern of gas streams flowing inwards and outwards  of the galactic halo. This situation can produce a complex scenario where different parts of a halo can be suffering opposite processes, with some regions being depleted of gas whereas the neighbour  regions  are being replenished with gas from outside of the halo. 
These results would establish a direct link between the turbulent ICM and the efficiency of the RPS effect. 
As it is known, the degree of turbulence of the ICM in clusters scales with the 
cluster total mass, being higher in massive clusters \citep[e.g.][]{Ryu_2008, Schmidt_2016}. 
Thus, galaxies in more massive clusters 
would experience an even more turbulent environment that would lead to a more dramatic reduction of the RPS effect compared with the traditional approach, where galaxies in massive clusters are more stripped due to their higher relative velocities and the denser ICM. 
In any case, it seems that  the scenario emerging from our results would be inconsistent with the traditional approaches to study the RPS where galaxies are usually affected by  homogeneous flows acting onto the galactic halo.

The existence of a complex pattern of flows, turbulence and a constant fueling of gas to the hot corona from the ICM  could produce, according to the results presented in this paper, that the global effect of the interaction of the galaxies with their environment could be substantially less dramatic than suggested in earlier works. In these references, it is mostly accepted that the hot corona can be quickly swept away in scales of few Myrs \citep[e.g.][]{2008MNRAS.383..593M,2015MNRAS.449.2312V} and that the cold component is also dramatically affected but on larger temporal scales \citep[e.g.][]{2000Sci...288.1617Q}. However, the results presented in this paper show that all galaxies retain an important fraction of bound gas, both hot and cold, being in all cases the hot bound component dominant. These striking results would be in agreement with  the observational results  \citep[e.g.][]{2007ApJ...671..190S, 2008ApJ...679.1162J, 2016ApJ...826..167G} evidencing that all observed galaxies exhibit a hot galactic corona. \cite{2010ApJ...715L...1M} suggested that the temperature gradient between the hotter gas in the ICM and the relatively cooler gas in the corona would translate into a pressure gradient competing with the RPS. The results presented in this paper could be explained by a combination of these two effects: on the one side, the confining action of the temperature gradient between the ICM and the cooler external part of the galactic halo and, on the other side,  the gas flows from the ICM that continuously percolate the corona and, therefore, reduce the relevance of the interaction with the environment. 

A final comment on the role of galaxy mergers on the results described in this paper has to be done. These events could modify the star 
formation rates and, consequently, the amount of cold gas. However, at the same time, the mergers also
act as a heating mechanism of the gaseous halo by transforming gravitational and kinetical energy into thermal energy of 
the gas in the galactic halo \citep[e.g.][]{2013MNRAS.436.3507N}. These two sides of the same process would compete removing and locking
cold gas, being quite unclear which would be the net effect. The results shown in Fig.~\ref{fig10} seem to ensure that, at least in our simulations, 
the RPS is the dominant effect.

The poor statistics of our numerical sample of galaxies, directly connected with the fact that our simulation describes simultaneously the whole computational volume without performing any re-simulation, does not allow us to draw stronger conclusions but present our result only as a global trend. A next generation of numerical simulations, already in project, could address this issue with higher accuracy.

\section*{Acknowledgements} 

 This work was  supported by the Spanish Ministerio de Econom\'{\i}a y Competitividad (MINECO)(grants  
AYA2013-48226-C03-02-P and AYA2016-77237-C3-3-P)   and    the   Generalitat   Valenciana   (grant  PROMETEOII/2014/069). Authors wish to thank Jorge S\'anchez, Claudio Dalla Vecchia, and Alexander Vazdekis for useful discussions, and the anonymous referee for his/her constructive criticism and enligthening comments. Simulations have been carried out in the supercomputer Lluis Vives  at the Servei d'Inform\`atica of the Univeristat de Val\`encia.

\bibliographystyle{mnras}
\bibliography{rps_6}

\end{document}